\documentclass{elsart}
\usepackage{amssymb,amsmath,epsfig}
\journal{Nuclear Physics B}
\def\beq{\begin{equation}}
\def\eeq{\end{equation}}
\def\beqa{\begin{eqnarray}}
\def\eeqa{\end{eqnarray}}

\begin{document}
\flushright{Preprint IFUP-TH/09-05}
\begin{frontmatter}

\title{Gravitational helicity interaction\thanksref{MIUR}}
\thanks[MIUR]{Research supported in part by M.I.U.R.}

%\author{A.\@ Barbieri\thanksref{abmail}}
\author{A.\@ Barbieri and E.\@ Guadagnini}
%\ead{andrea.barbieri@df.unipi.it}
%\author{E.\@ Guadagnini}%\corauthref{egmail}}
%\ead{enore.guadagnini@df.unipi.it}
\corauth{{\it Email addresses:} {\tt andrea.barbieri@df.unipi.it} ,\\ \indent 
{\tt enore.guadagnini@df.unipi.it} }
%\author{\and E.\@ Guadagnini\thanksref{egmail}}
%\thanks[egmail]{{\it Email address:} }pippo}
%\thanks[abmail]{{\it Email address:} {\tt andrea.barbieri@df.unipi.it}}
%\author{\and E.\@ Guadagnini\thanksref{egmail}}
%\thanks[egmail]{{\it Email address:} {\tt enore.guadagnini@df.unipi.it}}
\address{Dipartimento di Fisica {\sl Enrico Fermi} dell'Universit\`a di Pisa\\
and INFN Sezione di Pisa\\
Largo B.\@ Pontecorvo, 3. 56100-PISA-Italy}

\begin{abstract}
For gravitational deflections of massless particles of given helicity from a 
classical rotating body, we describe the general relativity corrections to the 
geometric optics approximation. We compute the corresponding scattering cross 
sections for neutrinos, photons and gravitons to lowest order in the 
gravitational coupling constant. We find that the helicity coupling to 
spacetime geometry modifies  the  ray deflection formula of the geometric 
optics, so that rays of different helicity are deflected by different amounts. 
We also discuss the validity range of the Born approximation. 
\end{abstract}

\end{frontmatter}

\section{Introduction}
Classical massive spinning (test) particles have been shown not to 
move in general along timelike geodesics, because the coupling 
of the particle spin to spacetime geometry produces, in the 
nonrelativistic limit,
both spin-orbit and spin-spin gravitational forces \cite{Mat,Papa,Wald72}. 
On the other hand, the massless limit of the Mathisson-Papapetrou equations
implies that classical massless spinning test 
particles follow null geodesics with the spin vector either 
parallel or antiparallel to their motion \cite{Ma75}. 
The early studies of the propagation of electromagnetic waves have indeed 
shown that  the geometric opticts limit predicts geodesic rays and parallel 
transport of the polarization vector 
\cite{Skrot,Balazs,Pleb}. The existence of a helicity asymmetry in the 
cross section for the scattering of electromagnetic waves 
off a Kerr black hole has been suggested, but it has not been explicitely 
computed, in \cite{MashPRD} and, by using dimensional  arguments, its 
magnitude has been estimated  in \cite{MashNat} 
for the deflection of light rays 
from the Sun in the $ \lambda \rightarrow 0$ approximation (as we shall see, 
in the radio waves region the Born approximation, pushed to the limit 
of its validity range, leads to a different prediction). 
The helicity asymmetry for electromagnetic 
waves has recently been explicitely computed, 
by resorting to quantum field theory methods in \cite{EG} 
and by using completely classical arguments in \cite{BG}. 

\begin{figure}\label{fig:waveScat}
\begin{center}
\epsfig{figure=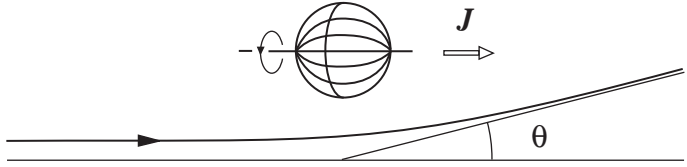, width=0.65\textwidth}
\end{center}
\caption{Deflection from a rotating body.}
\end{figure}

In this paper we extend the results of \cite{EG,BG} by computing, in the Born 
approximation, the scattering cross sections for fields of helicity 
$\pm \eta $ with $ \eta= 1/2 {\rm ~or~} 2$,  i.e. for neutrinos and 
gravitational waves\footnote{After completing this work, we became 
aware of the article \cite{De} in which, by using the Feynman-diagram 
technique, the  cross sections for the  gravitational  scattering of 
photons and gravitons from a classical rotating body have been computed. 
In the electromagnetic case, we find a complete agreement. However, for 
gravitational wave scattering, the classical calculation of the present 
paper leads to a different prediction. The correctness of our result has 
been checked  by using also the effective quantum field theory approach; 
we shall present the details  in a forthcoming paper.}. 
 In the case of a gravitational deflection from a 
classical rotating body with mass $M$ and angular momentum $J$ 
(directed along the incident wavevector, see Figure \ref{fig:waveScat}),  
in the $\theta\rightarrow 0$ limit, the asymmetry  turns out to be given by 
the following expression
\beq\label{eq:asymExpr}
\chi_\eta(\lambda,\theta)= \frac{(d\sigma_{+\eta}/d\Omega)-
(d\sigma_{-\eta}/d\Omega)}{(d\sigma_{+\eta}/d\Omega)
+(d\sigma_{-\eta}/d\Omega)}\simeq -\eta\frac{J}{Mc}\frac{2\pi}{\lambda}\theta^2
\eeq
where $d\sigma_{\pm\eta}/d\Omega$ is the differential cross section for the 
scattering of $\pm \eta$ helicity waves, $\theta$ is the scattering angle 
and $\lambda$ denotes the wavelength.  The validity of the Born approximation 
requires  $4\pi^2GJ\theta \ll c^3\lambda^2$.  The helicity asymmetry as 
expressed in equation (\ref{eq:asymExpr}) 
is actually meaningful only for wavelengths 
sufficiently large so that diffraction effects are important. When this is not 
the case, i.e. for wavelengths   $\lambda \ll 2\pi GM/c^2$,  the classical 
ray concept can meaningfully be used and  the cross sections imply that the 
relation between the impact parameter and the deflection angle is modified 
as follows
\beq\label{eq:deltaTheta}
\theta_{\pm \eta}(b)\simeq\theta_0(b)\left\{
1-\eta\left(\frac{1}{4}\pm \frac{J}{Mc}\frac{2\pi}{\lambda} \right)
\theta^2_0(b)|\ln \theta_0(b)|
+O(\theta_0(b)^3)
\right\}
\eeq
where $\theta_0(b)=4GM/(c^2b)$ is the standard result for the 
null geodesic deflection from a spherically symmetric body of mass $M$.
It should be noted  that expressions (\ref{eq:asymExpr}) and 
(\ref{eq:deltaTheta}) {\em do not} apply for $\lambda\rightarrow 0$; their 
domain of validity is discussed in section \ref{sec:discussion}.
We may thus conclude that, in analogy with the massive spinning (test) 
particle, massless spinning (test) particles do not move along null 
geodesics due to both a helicity-orbit coupling and to a helicity-spin 
coupling, corresponding respectively to the first and second term inside the 
round brackets in equation (\ref{eq:deltaTheta}).

The outline of the paper is as follows: in section \ref{sec:crossSecs} we 
compute the scattering cross sections in Born approximation for massless 
fields of helicity $\pm \eta=0,\pm 1/2,\pm 1,\pm 2$. In section 
\ref{sec:discussion} we derive equations (\ref{eq:asymExpr}) and 
(\ref{eq:deltaTheta}) and we discuss the range of application of these 
equations; finally, we give an estimate of the relevant effects 
in the case of deflections from the Sun and we discuss the case of large 
deflections in low frequency scattering off gravitationally compact objects.
In the following we adopt relativistic field theory units $c=1=\hbar$.

\section{Scattering cross sections}\label{sec:crossSecs}
\subsection{Scalar field}
In order to fix our notations and  conventions, we begin with the scalar 
massless field;  this will also be of help in separating the effects of 
``spin-orbit'' coupling from the helicity couplings
which we want to investigate in the following sections (the ``spin'' referred 
to is the angular momentum of the deflecting body).

To lowest order in the gravitational coupling constant, the metric tensor
at large distance from a localized stationary rotating body in an otherwise
empty spacetime can be put into the following form \cite{MTW}:
\beqa\label{eq:metric}
g_{\mu\nu}dx^\mu dx^\nu&=&-\left(1-\frac{2GM}{|\vec{x}|}\right)dt^2
+\left(1+\frac{2GM}{|\vec{x}|}\right)\delta_{ij}dx^i dx^j
\nonumber\\&&
-4\epsilon_{ijk}\frac{GJ^i x^j}{|\vec{x}|^3}dt\,dx^k
\eeqa
where multipoles of the body's stress-energy tensor higher than the second
are neglected, greek indices run from 0 to 3 with $t\equiv x^0$, 
latin indices run from 1 to 3
and $\epsilon_{ijk}$ is the completely antisymmetric tensor. $M$ and 
$\vec{J}\equiv M\vec{S}$ are respectively the total mass and angular momentum 
of the body. The minimally coupled equation of motion for a massless scalar 
field $\phi(x)$ 
in curved spacetime can be written as \cite{BD}
\beq\label{eq:fullS0}
0=\frac{1}{\sqrt{-|g|}}\partial_\mu \left(\sqrt{-|g|} g^{\mu\nu}\partial_\nu
\phi\right)
\eeq
where $|g|$ denotes the determinant of the covariant metric $g_{\mu\nu}$ 
and $g^{\mu\nu}$ is the inverse metric satisfying 
$g^{\mu\rho}g_{\rho\nu}=\delta^\mu_\nu$. In the same approximations of 
equation (\ref{eq:metric}), equation 
(\ref{eq:fullS0}) becomes
\beq\label{eq:redS0}
-\eta^{\mu\nu}\partial_\mu\partial_\nu\phi=-\left(h^{\mu\nu}
-\frac{1}{2}h^\rho{}_\rho\eta^{\mu\nu}\right)\partial_\mu\partial_\nu\phi
\eeq
where $\nobreak{h_{\mu\nu}=g_{\mu\nu}-\eta_{\mu\nu}}$ and 
$\eta_{\mu\nu}=\eta^{\mu\nu}={\rm diag}(-1,1,1,1)$ denotes the flat Minkowski 
metric, which is used to raise and lower all indices. If one puts
\beq
\phi(x)=e^{ik_\mu x^\mu}\phi^{{\rm (in)}}+
e^{-i\omega t}\phi^{{\rm (diff)}}(\vec{x})+\rm{c.c.}
\eeq
with $k^\mu k_\mu =0,\,k^0=\omega$, 
to lowest order in $G$ (Born approximation) the field 
$\phi^{{\rm (diff)}}$ satisfies the Helmoltz equation with a source term
\beq
-(\omega^2+\triangle)\phi^{{\rm (diff)}}(\vec{x})=
k^\mu k^\nu h_{\mu\nu}(\vec{x})
\, e^{i\vec{k}\cdot\vec{x}}\phi^{{\rm (in)}}\equiv F(\vec{k},\vec{x})
\eeq
In the $|\vec{x}|\rightarrow \infty$ limit, this equation is 
asymptotically solved by
\cite{JDJ}
\beq
\phi^{{\rm (diff)}}(\vec{x})\simeq\frac{e^{i\omega|\vec{x}|}}{|\vec{x}|}
\int d^3y\, e^{-i\vec{p}\cdot\vec{y}}\frac{F(\vec{k},\vec{y})}{4\pi}
\equiv \frac{e^{i\omega|\vec{x}|}}{|\vec{x}|}{\mathcal A}(\vec{k},\vec{p})
\eeq
where $\vec{p}=\omega\vec{x}/|\vec{x}|$ is the asymptotic wave vector
and the cross section for the scattering process $\vec{k}\rightarrow \vec{p}$
($\vec{k}$ is the incoming momentum, $\vec{p}$ the outgoing one) is given by
\beq\label{eq:csS0}
\frac{d\sigma_{\eta=0}}{d\Omega}=
\left|\frac{{\mathcal A}}{\phi^{\rm (in)}}\right|^2
\eeq

By putting all together one finds 
\beq\label{eq:ampliS0}
{\mathcal A}(\vec{k},\vec{p})=\frac{GM}{\sin^2(\theta/2)}
\left[1+\frac{i}{\omega}\vec{S}\cdot\vec{p}\times\vec{k}\right]
\phi^{{\rm (in)}}
\eeq

\subsection{Neutrinos}
Neutrinos are described in flat spacetime by massless fields transforming 
according to the $(1/2,0)$ or $(0,1/2)$ 
representations of $SL(2,\Cset)$ and their 
coupling to spacetime geometry 
is accomplished by introducing an orthonormal basis $\{\Theta_I^\mu\}$ on the 
tangent space, usually called tetrad or vierbein, so that \cite{BD}
\beq\label{eq:defVierbein}
g_{\mu\nu}(x)\Theta_I^\mu(x)\Theta_J^\nu(x)=\eta_{IJ}
\eeq
where uppercase latin indices run from 0 to 3 and denote internal Lorentz 
indices. The spin connection is an $SL(2,\Cset)$-valued connection defined by 
\beq\label{eq:connDer}
\Omega^{IJ}_\mu\equiv \frac{1}{2}\left(E^I_\rho \nabla_\mu \Theta^{J\rho}
-E^J_\rho \nabla_\mu \Theta^{I\rho}\right)
\eeq
where $\{E^I_\mu\}$ is the cotetrad field satisfying 
\beq
E^I_\mu(x)\Theta_J^\mu(x)=\delta^I{}_J
\eeq
and Lorentz indices are raised an lowered with the Minkowski metric
$\eta_{IJ}$. Let  $L$ be a $SL(2,\Cset)$-valued gauge transformation
\beq
L:M\rightarrow SL(2,\Cset), x\mapsto L(x) 
\eeq
and let  $[L_{(r)}]$ denote the matrix representing $L$ in the 
representation $(r)$; the tetrad is
defined up to
\beq
\Theta_I^\mu(x)\rightarrow [L_{(\frac{1}{2},\frac{1}{2})}(x)]^{I'}{}_I
\Theta_{I'}^\mu(x)
\eeq
so that, if the field $\psi^a_{(r)}(x)$ transforms as
\beq\label{eq:trasField}
\psi_{(r)}(x)\rightarrow [L_{(r)}(x)]^a{}_{a'} \psi^{a'}_{(r)}(x)
\eeq
and transforms as a scalar field under a change of coordinates, 
the derivative
\beq\label{eq:covDer}
D_\mu \psi_{(r)}^a\equiv \partial_\mu \psi_{(r)}^a
- \frac{i}{2}\Omega_{IJ\mu}[J_{(r)}^{IJ}]^a{}_{a'}\psi_{(r)}^{a'}
\eeq
transforms covariantly. The generators $[J_{(r)}^{IJ}]^a{}_{a'}$ 
of $SL(2,\Cset)$ in the representation $(r)$ satisfy the commutation relations
\beq
[J_{(r)}^{IJ},J_{(r)}^{HK}]=i\left(\eta^{IH}[J_{(r)}^{JK}]
+\eta^{JK}[J_{(r)}^{IH}]-\eta^{IK}[J_{(r)}^{JH}]-\eta^{JH}[J_{(r)}^{IK}]\right)
\eeq

The propagation equation for neutrinos with helicity $\pm 1/2$ 
in flat spacetime can be written as
\beq\label{eq:flatEqns}
[\Gamma_{(\pm)}^I]^a{}_{a'} \partial_I \psi_{(\pm)}^{a'}=0
\eeq
where $a$ is spinor index and 
\beq\label{eq:defGamma}
[\Gamma_{(\pm)}^I]^a{}_{a'}=
\begin{cases}
\delta^a{}_{a'} & {\rm for~} I=0 \cr & \cr
\pm [\sigma^I]^a{}_{a'} & {\rm for~} I\neq 0
\end{cases}
\eeq
with $[\vec{\sigma}]^a{}_{a'}$ the Pauli spin matrices; 
the $SL(2,\Cset)$ generators are given by
\beq
[J_{(\pm)}^{IJ}]^a{}_{a'}=
\begin{cases}
\pm {\displaystyle \frac{i}{2}} 
[\sigma^J]^a{}_{a'} & {\rm for~} I=0,\, J\neq 0\cr &\cr
{\displaystyle \frac{\epsilon_0{}^{IJ}{}_K}{2}}
[\sigma^K]^a{}_{a'} & {\rm for~} I,J\neq 0
\end{cases}
\eeq
with $\epsilon_{0123}=1$. The covariant field equations in a non flat 
spacetime are thus obtained by substituting the coordinate derivative 
$\partial_I$ in equation (\ref{eq:flatEqns}) with the covariant derivative
$D_I\equiv \Theta^\mu_I D_\mu$ of equation (\ref{eq:covDer}) \cite{BD}:
\beq\label{eq:curvedEqns}
[\Gamma_{(\pm)}^I]^a{}_{a'} D_I \psi_{(\pm)}^{a'}=0
\eeq
To first order in $G$, the tetrad and cotetrad fields can be put in the form
\beqa
\Theta_0^\mu &=& \left(1+\frac{GM}{|\vec{x}|}\right)\delta_0^\mu
+\epsilon_{ijk}\frac{GJ^j x^k}{|\vec{x}|^3}\delta_i^\mu
\nonumber\\
\Theta_i^\mu &=& \left(1-\frac{GM}{|\vec{x}|}\right)\delta_i^\mu
-\epsilon_{ijk}\frac{GJ^j x^k}{|\vec{x}|^3}\delta_0^\mu
\nonumber\\
E^0_\mu &=& \left(1-\frac{GM}{|\vec{x}|}\right)\delta^0_\mu
+\epsilon_{ijk}\frac{GJ^j x^k}{|\vec{x}|^3}\delta^i_\mu
\nonumber\\
E^i_\mu &=& \left(1+\frac{GM}{|\vec{x}|}\right)\delta^i_\mu
-\epsilon_{ijk}\frac{GJ^j x^k}{|\vec{x}|^3}\delta^0_\mu
\nonumber
\eeqa
so that the spin connection $\Omega^{IJ}_\mu$ is given by
\beqa
\Omega^{0i}_\mu &=& -\partial_i\left(\frac{GM}{|\vec{x}|}\right)\delta^0_\mu
+\partial_i\left(\frac{\epsilon_{jhk}GJ^h x^k}{|\vec{x}|^3}\right)\delta^j_\mu
\nonumber\\
\Omega^{ij}_\mu &=& -\partial_i\left(\frac{GM}{|\vec{x}|}\right)\delta^j_\mu
+\partial_i\left(\frac{\epsilon_{jhk}GJ^h x^k}{|\vec{x}|^3}\right)\delta^0_\mu
- (i\leftrightarrow j)
\nonumber
\eeqa
and the propagation equation (\ref{eq:curvedEqns}) becomes
\beqa
0&=&\partial_t \psi_{(\pm)}\pm [\vec{\sigma}]\cdot\vec{\nabla}
\psi_{(\pm)}
-\frac{G}{2}\vec{\nabla}\cdot\left(\frac{\vec{J}\times\vec{x}}{|\vec{x}|^3}
\pm \frac{M\left[\vec{\sigma}\right]}{|\vec{x}|}
\right)\psi_{(\pm)}
\nonumber\\
&&
+G\left(\frac{M}{|\vec{x}|}
\mp \frac{ \vec{J} \times \vec{x}\cdot
[\vec{\sigma}]}{|\vec{x}|^3}\right)
\partial_t \psi_{(\pm)}
+G\left(\frac{\vec{J}\times\vec{x}}{|\vec{x}|^3}
\mp M \frac{[\vec{\sigma}]}{|\vec{x}|}\right)
\cdot\vec{\nabla} \psi_{(\pm)}
\label{eq:diffusion}
\eeqa
Again, the field $\psi_{(\pm)}(x)$ can be written as the sum of 
an incident and a diffracted wave component
\beq
\psi_{(\pm)}(x)=e^{-i\omega t}\left[e^{i\vec{k}\cdot\vec{x}}
\psi^{\rm (in)}_{(\pm)}
+\psi^{\rm (diff)}_{(\pm)}(\vec{x})\right]
\eeq
where $|\vec{k}|=\omega$, and $\psi^{\rm (in)}_{(\pm)}$ satisfies
\beq\label{incSwave}
\vec{k}\cdot\left[\right.\!\vec{\sigma}\!\left.\right]
\psi^{\rm (in)}_{(\pm)}=\pm\omega\,
\psi^{\rm (in)}_{(\pm)}
\eeq
To first order in the gravitational constant the diffracted 
component satisfies the equation
\beq\label{eq:IstOrder}
-i\omega \psi^{\rm (diff)}_{(\pm)}\pm[\vec{\sigma}]
\cdot\vec{\nabla} \psi^{\rm (diff)}_{(\pm)}=
F_{(\pm)}(\vec{k},\psi^{\rm (in)}_{(\pm)},\vec{x})
\eeq
with
\beqa
F_{(\pm)}(\vec{k},\psi^{\rm (in)}_{(\pm)},\vec{x})&=&
iGM \omega \left[\frac{2}{|\vec{x}|}
\mp\frac{ \vec{S} \times \vec{x}}{|\vec{x}|^3}\cdot\left(
[\vec{\sigma}]\pm \frac{\vec{k}}{\omega}\right)\right]
e^{i\vec{k}\cdot\vec{x}}\psi^{\rm (in)}_{(\pm)}
\nonumber\\
&&
+\frac{GM}{2}\vec{\nabla}\cdot\left(\frac{\vec{S}\times\vec{x}}{|\vec{x}|^3}
\pm \frac{\left[\vec{\sigma}\right]}{|\vec{x}|}
\right)
e^{i\vec{k}\cdot\vec{x}}\psi^{\rm (in)}_{(\pm)}
\eeqa
By left multiplying equation (\ref{eq:IstOrder}) with the operator
$i\omega \pm [\vec{\sigma}]\cdot\vec{\nabla}$ one obtains
\beq\label{eq:helmoltz}
\left(\omega^2+\nabla^2\right) \psi^{\rm (diff)}_{(\pm)}=
\left(i\omega  \pm [\vec{\sigma}]\cdot\vec{\nabla}\right)F_{(\pm)}
\eeq
The asymptotic $|\vec{x}|\rightarrow\infty$ solution to
equation (\ref{eq:helmoltz}) is 
\beqa
\psi^{\rm (diff)}_{(\pm)}(\vec{x})&\simeq&
-\frac{e^{i\omega|\vec{x}|}}{|\vec{x}|}\int d^3y
\frac{e^{-i\vec{p}\cdot\vec{y}}}{4\pi}
\left(i\omega  \pm [\vec{\sigma}]\cdot\vec{\nabla}\right)
F_{(\pm)}(\vec{k},\psi^{\rm (in)}_{(\pm)},\vec{y})
\nonumber\\
&=& -i\frac{e^{i\omega|\vec{x}|}}{|\vec{x}|}
\left(\omega  \pm \vec{p}\cdot[\vec{\sigma}]\right)\int d^3y
\frac{e^{-i\vec{p}\cdot\vec{y}}}{4\pi}
F_{(\pm)}(\vec{k},\psi^{\rm (in)}_{(\pm)},\vec{y})
\nonumber\\
&\equiv&
-i\frac{e^{i\omega|\vec{x}|}}{|\vec{x}|}
\left(\omega  \pm \vec{p}\cdot[\vec{\sigma}]\right)
{\mathcal A}_{(\pm)}(\vec{k},\psi^{\rm (in)}_{(\pm)},\vec{p})
\eeqa
where $\vec{p}=\omega\vec{x}/|\vec{x}|$ is the outgoing momentum. The
differential cross section for the scattering process
$\vec{k}\rightarrow\vec{p}$ is thus given by
\beq\label{eq:csS1/2}
\frac{d\sigma_{\eta=\pm 1/2}}{d\Omega}={|(\psi^{\rm out}_{(\pm)})^\dagger
{\mathcal A}_{(\pm)}|^2}
\eeq
with $\psi^{\rm out}_{(\pm)}$ a constant spinor satisfying
\beq\label{outSwave}
\vec{p}\cdot[\vec{\sigma}]\psi^{\rm out}_{(\pm)}=\pm \omega\,
\psi^{\rm out}_{(\pm)} \quad ,\hskip 0.6cm
(\psi^{\rm out}_{(\pm)})^\dagger \psi^{\rm out}_{(\pm)}=
\frac{4 \omega^2}{(\psi^{\rm (in)}_{(\pm)})^\dagger 
\psi^{\rm (in)}_{(\pm)}}
\eeq
Up to an irrelevant phase, equations (\ref{incSwave}) and (\ref{outSwave})
imply
\beqa
(\psi^{\rm out}_{(\pm)})^\dagger \psi^{\rm (in)}_{(\pm)} &=&
-2i\omega\cos(\theta/2)
\\
(\psi^{\rm out}_{(\pm)})^\dagger \left[\vec{\sigma}\right]
\psi^{\rm (in)}_{(\pm)}&=&-\frac{i}{\omega\cos(\theta/2)}\left[
i\vec{p}\times\vec{k}\pm\omega(\vec{p}+\vec{k}) 
\right]
\eeqa
so that a straightforward computation yields
\beqa
(\psi^{\rm out}_{(\pm)})^\dagger{\mathcal A}_{(\pm)}
&=&
\frac{GM}{\sin^2(\theta/2)}\cos(\theta/2)
\left[1+\frac{i}{2\omega}(\vec{S}\cdot\vec{p}\times\vec{k})
\left(1+\frac{1}{\cos^2(\theta/2)}\right)
\right.\nonumber\\&&\hskip 3.5cm \left.
\mp \frac{1}{2}\tan^2(\theta/2)\vec{S}\cdot(\vec{k}+\vec{p})\right]
\label{eq:ampliS1/2}
\eeqa

\subsection{Electromagnetic waves}
The scattering cross section for electromagnetic waves has been computed 
in \cite{EG,BG} and we simply quote the result. For the process 
$(\vec{k},\vec{\epsilon}^{\rm \,(in)})\rightarrow 
(\vec{p},\vec{\epsilon}^{\rm \,(out)})$ one obtains
\beq
\frac{d\sigma_{\eta=1}}{d\Omega}=\frac{
|{\mathcal A}(\vec{k},\vec{\epsilon}^{\rm \,(in)},
\vec{p},\vec{\epsilon}^{\rm \,(out)})|^2
}{(\vec{\epsilon}^{\rm \,(out)})^*\cdot\vec{\epsilon}^{\rm \,(out)}
(\vec{\epsilon}^{\rm \,(in)})^*\cdot\vec{\epsilon}^{\rm \,(in)}
}
\label{eq:csS1}
\eeq
with
\beqa
{\mathcal A}(\vec{k},\vec{\epsilon}^{\rm \,(in)},
\vec{p},\vec{\epsilon}^{\rm \,(out)})&=&\frac{GM}{\sin^2(\theta/2)}
\left\{
\left(\frac{1+\cos\theta}{2}+\frac{i}{\omega}\vec{S}\cdot\vec{p}\times\vec{k}
\right)(\vec{\epsilon}^{\rm \,(out)})^*\cdot\vec{\epsilon}^{\rm \,(in)}
\right.\nonumber\\&&\hskip -1cm \left.
-\frac{1}{2\omega^2}
(\epsilon^{\rm \,(out)}_i)^*\epsilon^{\rm \,(in)}_j
\left[k^i p^j
+i\omega S^h q^k\left(\epsilon_{hki} p^j+ \epsilon_{hkj} k^i\right)
\right]
\right\}
\label{eq:ampliS1}
\eeqa
where $ \vec q = \vec p - \vec k $. 

\subsection{Gravitational waves}
The lowest order scattering of gravitational waves can likewise be computed by 
means of an expansion in powers of the gravitational constant, but since 
the scattering is actually induced by the self-interactions of the 
gravitational field, 
one needs to be more careful in writing the corresponding equations.

One can power expand the metric tensor and the energy momentum tensor as
\beqa
g_{\mu\nu}(x)&=&\eta_{\mu\nu}+\ell \sum_{n=0}^\infty
h^{(n)}_{\mu\nu}(x)\times \ell^n
\label{expMetric}\\
T_{\mu\nu}(x)&=& \sum_{n=0}^\infty T^{(n)}_{\mu\nu}(x)\times \ell^n
\label{expTtens}
\eeqa
where $\ell=\sqrt{G}$  (the extra $\ell$ factor in equation 
(\ref{expMetric}) has been introduced in order to give $h^{(0)}_{\mu\nu}$ 
the canonical dimensions of a free field). The Einstein equations
\beq
G_{\mu\nu}(g)=8\pi \ell^2 T_{\mu\nu}
\eeq
are likewise expanded, up to and including terms of order $\ell^2$, as
\beqa
{\mathcal G}[h^{(0)}]&=&0
\label{freeGrav}\\
{\mathcal G}[h^{(1)}]+{\mathcal G}'[h^{(0)},h^{(0)}]
&=&8\pi T^{(0)}
\label{coulGrav}\\
{\mathcal G}[h^{(2)}]+2{\mathcal G}'[h^{(1)},h^{(0)}]
+{\mathcal G}''[h^{(0)},h^{(0)},h^{(0)}]&=&8\pi T^{(1)}
\label{deflGrav}
\eeqa
where ${\mathcal G}$ is the wave operator for a massless spin-2 field in
Minkowski space:
$$
{\mathcal G}[h]_{\mu\nu}=\frac{1}{2}\left[-h_{\mu\nu,\alpha}{}^\alpha
-h^\alpha{}_{\alpha,\mu\nu}+h_{\alpha\mu,\nu}{}^\alpha
+h_{\alpha\nu,\mu}{}^\alpha
-\eta_{\mu\nu}\left(h^{\alpha\beta}{}_{,\alpha\beta}
-h^\alpha{}_{\alpha,\beta}{}^\beta
\right)\right]
$$
Equations (\ref{freeGrav}-\ref{deflGrav}) admit a simple interpretation. 
Equation
(\ref{freeGrav}) describes free gravitational waves which propagate in flat
Minkowski space and which encode the perturbative degrees of
freedom of the gravitational field. Equation (\ref{coulGrav}) describes the
linearized gravitational field due to the energy momentum distribution of
matter and  of free gravitational waves. Note that, to this order,
the energy momentum tensor $T^{(0)}_{\mu\nu}$ must satisfy the flat space
conservation law $T^{(0)\mu\nu}{}_{,\nu}=0$. Finally, equation
(\ref{deflGrav}) describes the lowest order correction to the gravitational
wave propagation due to the linearized background as well as the backreaction
on the matter distribution of the zeroth order gravitational wave, because
$T^{(1)}$ must satisfy
\beq
T^{(1)\mu\nu}{}_{,\nu}=-T^{(0)\mu\lambda} \Gamma^{(0)\nu}{}_{\nu\lambda}
-T^{(0)\lambda\nu}\Gamma^{(0)\mu}{}_{\nu\lambda}
\label{corrTtens}
\eeq

In the limit of {\em weak} incoming gravitational waves, 
 in equations (\ref{coulGrav}) and (\ref{deflGrav}) one can
neglect the terms which
are non linear in $h^{(0)}$, so that, in order to obtain the lowest order
scattering, we only need to compute ${\mathcal G}'$, which is given by 
\cite{MTW}
\beqa
{\mathcal G}'[h,s]_{\mu\nu}&=&
-\frac{1}{4}\left(\eta_{\mu\nu}h^{\alpha\beta}
+h_{\mu\nu}\eta^{\alpha\beta}\right){\mathcal R}[s]_{\alpha\beta}
-\frac{1}{4}\left(\eta_{\mu\nu}s^{\alpha\beta}
+s_{\mu\nu}\eta^{\alpha\beta}\right){\mathcal R}[h]_{\alpha\beta}
\nonumber\\
&&+\left(\delta^\alpha_\mu\delta^\beta_\nu
-\frac{1}{2}\eta_{\mu\nu}\eta^{\alpha\beta}\right)
{\mathcal R}'[h,s]_{\alpha\beta}
\eeqa
with
\beqa
{\mathcal R}[h]_{\mu\nu}&=&\frac{1}{2}\left(-h_{\mu\nu,\alpha}{}^\alpha
-h^\alpha{}_{\alpha,\mu\nu}+h_{\alpha\mu,\nu}{}^\alpha
+h_{\alpha\nu,\mu}{}^\alpha\right)
\nonumber\\
{\mathcal R}'[h,s]_{\mu\nu}&=&\frac{1}{4}\left[
\frac{1}{2}h_{\alpha\beta,\mu}s^{\alpha\beta}{}_{,\nu}
+h^{\alpha\beta}\left(s_{\alpha\beta,\mu\nu}+s_{\mu\nu,\alpha\beta}
-s_{\alpha\mu,\beta\nu}-s_{\alpha\nu,\beta\mu}\right)
\right.\nonumber\\&&\hskip -1cm \left.
+h_\nu{}^{\alpha,\beta}\left(s_{\alpha\mu,\beta}-s_{\beta\mu,\alpha}\right)
-\left(h^{\alpha\beta}{}_{,\beta}-\frac{1}{2}h_\beta{}^{\beta,\alpha}\right)
\left(s_{\alpha\mu,\nu}+s_{\alpha\nu,\mu}-s_{\mu\nu,\alpha}\right)
\right]
\nonumber\\&&\hskip -1cm
+h\leftrightarrow s
\eeqa
Equations (\ref{freeGrav}-\ref{deflGrav}) become
\beqa
{\mathcal G}[h^{(0)}]&=&0
\label{freeGravLin}\\
{\mathcal G}[h^{(1)}]&=&8\pi T^{(0)}
\label{coulGravLin}\\
{\mathcal G}[h^{(2)}]&=&8\pi T^{(1)}-2{\mathcal G}'[h^{(1)},h^{(0)}]
\label{deflGravLin}
\eeqa

Let the incident wave be a transverse-traceless plane wave
\beq
h^{(0)}_{\mu\nu}(x)=\pi^{\rm (in)}_{\mu\nu}
{\rm e}^{{\rm i} k^\mu x_\mu}
+\rm{c.c.}
\eeq
with $k^\mu k_\mu = \pi^{{\rm (in)}\mu}{}_\mu = 0 =
k^\mu \pi^{\rm (in)}_{\mu\nu}$. Since 
$T^{(0)}_{\mu\nu}$ is assumed to be  stationary, one can choose 
an adapted coordinate system
$\{x^\mu\}=\{t,\vec{x}\}$ in which $T^{(0)}_{\mu\nu,0}=0$ and the linearized
Coulomb metric $h^{(1)}$ is given by
\beq
h^{(1)}_{\mu\nu}(\vec{x})=4\left(\delta^\alpha_\mu\delta^\beta_\nu-
\frac{1}{2}\eta_{\mu\nu}\eta^{\alpha\beta}\right)\int
\frac{T^{(0)}_{\alpha\beta}(\vec{x}')}{|\vec{x}-\vec{x}'|}d^3 x'
\eeq
For a pointlike spinning 
body placed at rest on the origin of spatial coordinate,
the zeroth order energy momentum tensor is 
\beq
T^{(0)}_{\mu\nu}(t,\vec{x})=M\delta_\mu^0\delta_\nu^0\delta^3(\vec{x})
-\frac{1}{2}(\delta_\mu^0\delta_\nu^i+\delta_\nu^0\delta_\mu^i)
\epsilon_{ijk}J^j\partial_k\delta^3(\vec{x})
\eeq
and the metric $\eta_{\mu\nu}+G h^{(1)}_{\mu\nu}(\vec{x})$ is precisely the 
metric of equation (\ref{eq:metric}). Equation (\ref{corrTtens}) is now 
solved by
\beq
T^{(1)}_{\mu\nu}=\left(\eta_{\nu\rho}\partial_\mu
+\eta_{\mu\rho}\partial_\nu-\eta_{\mu\nu}\partial_\rho\right)
\int d^4 x' \Gamma^{(0)\rho}{}_{\alpha\beta}(x')T^{(0)\alpha\beta}(x')
G(x,x')
\label{solCorrTtens}
\eeq
with $G(x,x')$ is a Green function for the 4-dimensional scalar wave
operator (the ambiguity related to the particular choice of the Green function 
does not affect the scattering cross
section of transverse traceless waves). Given the harmonic time dependence 
of $h^{(0)}$ and the stationarity of
$T^{(0)}$, one can set 
\beq
h^{(2)}_{\mu\nu}=\left(\delta^\alpha_\mu\delta^\beta_\nu-
\frac{1}{2}\eta_{\mu\nu}\eta^{\alpha\beta}\right)
w_{\alpha\beta}(\vec{x}){\rm e}^{-{\rm i}\omega t}
+\rm{c.c.}
\eeq
with $\omega=k^0$ and $w_{0\mu}=0$. With this choice, the $\mu=0=\nu$ 
and $\mu=0,\,\nu=i$ components of equation
(\ref{deflGravLin}) yield respectively
\beqa
w_{ii}(\vec{x})&=&-2\frac{D_{00}(\vec{x})}{\omega^2}\\
w_{ij,j}(\vec{x})&=& 2{\rm i}\frac{D_{0i}(\vec{x})}{\omega}
-2\frac{D_{00,i}(\vec{x})}{\omega^2}
\eeqa
where 
\beq
D_{\mu\nu}=-\left(8\pi T^{(1)}_{\mu\nu}-
2{\mathcal G}'[h^{(1)},h^{(0)}]_{\mu\nu}\right)
{\rm e}^{{\rm i}\omega t}
\eeq

Finally, the space-space components of equation (\ref{deflGravLin}) yield
the relevant equation for the scattering,
\beq
(\omega^2+\triangle)w_{ij}(\vec{x})=2\left(D_{ij}-\frac{D_{00,ij}}{\omega^2}
+{\rm i}\frac{D_{0i,j}+D_{0j,i}}{\omega}\right)\equiv 
F_{ij}(\vec{k},\pi^{\rm (in)},\vec{x})
\eeq
whose asymptotic $|\vec{x}|\rightarrow\infty$ solution is given by
\beq
w_{ij}(\vec{x})\simeq\frac{{\rm e}^{{\rm i} \omega |\vec{x}|}}{|\vec{x}|}
\int d^3\! y\, {\rm e}^{-{\rm i}\vec{p}\cdot\vec{y}}
\left(-\frac{F_{ij}(\vec{k},\pi^{\rm (in)},\vec{y})}{4\pi}\right)
\equiv \frac{{\rm e}^{{\rm i} \omega |\vec{x}|}}{|\vec{x}|} 
{\mathcal A}_{ij}(\vec{k},\pi^{\rm (in)},\vec{p})
\eeq
with $\vec{p}=\omega \vec{x}/|\vec{x}|$.

The cross section for the gravitational scattering into gravitons with 
wavevector $\vec{p}$ and polarization $\pi^{{\rm (out)}}_{ij}$, with
$p_i\pi^{{\rm (out)}}_{ij}=0=\pi^{{\rm (out)}}_{ii}$,
is given by \cite{Wein}
\beq
\frac{d\sigma_{\eta=2}}{d\Omega}=G^2
\frac{\left|(\pi^{{\rm (out)}}_{ij})^* {\mathcal A}_{ij}\right|^2}{
(\pi^{{\rm (out)}}_{ij})^* \pi^{{\rm (out)}}_{ij}
(\pi^{{\rm (in)}}_{hk})^* \pi^{{\rm (in)}}_{hk}}
\label{eq:csS2}
\eeq
This expression shows that terms in $F_{ij}$ which either are total 
derivatives or 
are proportional to the metric tensor, such as $T^{(1)}$, see equation 
(\ref{solCorrTtens}),
do not contribute to $ d \sigma_{\eta=2} $
and need not be computed at all. Therefore
\beq
F_{ij}\sim \left(4{\mathcal R}'[h^{(1)},h^{(0)}]_{ij}
-h^{(0)}_{ij}\eta^{\alpha\beta}{\mathcal R}[h^{(1)}]_{\alpha\beta}\right)
{\rm e}^{{\rm i}\omega t}
\eeq
and
\beqa
\pi^{{\rm (out)}*}_{ij} {\mathcal A}_{ij}&=&\frac{M}{\sin^2(\theta/2)}
\left[(\pi^{{\rm (out)}*}_{ij} \pi^{{\rm (in)}}_{ij})
(1+\frac{{\rm i}}{\omega}\vec{S}\cdot\vec{p}\times\vec{k})
\right.\nonumber\\&&\hskip 1cm\left.
+\frac{{\rm i}}{\omega} (\pi^{{\rm (out)}*}_{ih}\pi^{{\rm (in)}}_{hj})
(q_i\epsilon_{jpq} - q_j\epsilon_{ipq})S_p q_q
\right]
\label{eq:ampliS2}
\eeqa
where $ \, \vec q = \vec p - \vec k $. 

\section{Gravitational helicity interaction}\label{sec:discussion}
\subsection{The helicity asymmetry}
By rotating the spatial coordinates so that 
$\vec{k}=\omega\hat{z}$ and $\vec{p}=\omega(\cos\theta \hat{z}
+\sin\theta\hat{x})$, ($\theta >0$ denotes the scattering angle) and  
by writing the amplitudes (\ref{eq:ampliS1}) and (\ref{eq:ampliS2}) 
in matrix form with respect to circular polarization states for the incident 
and scattered waves, one finds that helicity 
changing terms are exactly zero for $\eta=1/2$ and $\eta= 1$, 
while for $\eta=2$ they are $O(\theta^4)$ with respect to the Rutherford 
amplitude; this is not unexpected, because, in the 
language of Feynman diagrams, 
the amplitude for scattering of gravitons off a massive (elementary) 
particle contains, besides the one-graviton-exchange graph which conserves 
helicity, also the ``Compton'' graph where the external gravitons couple 
directly to the massive particle (the total amplitude factorizes nicely, 
see \cite{ChoiEtAl}).
The helicity conserving terms, on the other hand, are in general different 
from each other and in the $\theta\rightarrow 0$ limit the cross section for 
the scattering of $\pm\eta$ helicity waves is
\beqa\label{eq:helAmpli}
\frac{d \sigma_{\pm\eta}}{d \Omega}(\vec{k},\vec{p})
=\frac{G^2M^2}{\sin^4(\theta/2)}
\left|1+i\omega S_y \theta-\frac{\eta}{2}
\left(\frac{1}{4} \pm \omega S_z\right)\theta^2
+O(\theta^3)
\right|^2
\eeqa
Expression (\ref{eq:helAmpli}) is valid for $\eta = 0, 1/2 , 1, 2 $ and 
summarizes the results contained in equations (\ref{eq:csS0}), 
(\ref{eq:ampliS0}), (\ref{eq:csS1/2}), (\ref{eq:ampliS1/2}), (\ref{eq:csS1}), 
(\ref{eq:ampliS1}), (\ref{eq:csS2}), (\ref{eq:ampliS2}).  
The meaning of the various terms, besides the usual 
Coulomb one, appearing in the right hand side of equation (\ref{eq:helAmpli}) 
can be understood as follows. 
The imaginary term $i\omega S_y \theta $ represents a ``spin-orbit'' 
interaction;  it does not depend on the helicity of the scattered field and it 
merely couples  the body's angular momentum component which is orthogonal to 
the scattering plane to the orbital motion of the waves. This component is 
also present for scalar fields. 
The term  $ - \eta \theta / 8$  does not depend on the sign of helicity and 
can be described as a ``helicity-orbit'' interaction because it is independent 
of the chirality of the wave and of the body's angular momentum.  
Finally, the term $\mp \eta \omega S_z / 2$  describes the effects of the 
``helicity-spin'' interaction, which distinguishes the two chiralities.

The helicity asymmetry $\chi(\theta)$ at a deflection angle 
$\theta$ (such that  $\omega S_z\theta^2 \ll 1$) is thus
\beq
\chi_\eta(\lambda,\theta)= \frac{(d\sigma_{+\eta}/d\Omega)-
(d\sigma_{-\eta}/d\Omega)}{(d\sigma_{+\eta}/d\Omega)
+(d\sigma_{-\eta}/d\Omega)}\simeq -\eta\omega S_z \theta^2
\eeq
which coincides with equation (\ref{eq:asymExpr}) and shows that a linearly 
polarized or unpolarized incident plane wave is detected 
at the deflection angle $\theta$ with a net nontrivial 
left circular polarization.

\subsection{Orbit corrections}
Let us now assume for simplicity that the incoming wave vector $\vec{k}$ 
is aligned with the angular momentum vector $\vec{J}$, as shown in 
Figure \ref{fig:waveScat}; 
i.e. $\vec{k}\times\vec{J}=0$ and 
$\vec{k}\cdot\vec{J}= | \vec{k} | |\vec{J}| $. 
If diffraction  effects are negligible, the semiclassical description 
of the scattering is valid. 
In this case, the $\theta\rightarrow 0$ limit of wave scattering admits  
a description in 
terms of particle orbits, where the dependence of the deflection angle 
$\theta$ on the impact parameter is given by \cite{LL}
\beq\label{eq:defImpact}
\frac{d\sigma}{d\Omega}(\theta)=\frac{b(\theta)}{\sin \theta}
\left|\frac{d b(\theta)}{d\theta}\right|
\eeq
By substituting equation (\ref{eq:helAmpli}) into equation 
(\ref{eq:defImpact}) one finds
\beq
b_{\pm\eta}(\theta)=\frac{4GM}{\theta}\left\{
1-\eta\left(\frac{1}{4}\pm\omega S_z\right)\theta^2 |\ln\theta |
+O(\theta^3)
\right\}
\eeq
leading to expression (\ref{eq:deltaTheta}) for the deflection angle 
$\theta_{\pm \eta}(b)$ at fixed impact parameter $b$.

\subsection{Domain of validity}
We now discuss the range of application of equations (\ref{eq:helAmpli}),
(\ref{eq:asymExpr}) and (\ref{eq:deltaTheta}), which represent the main 
results 
of this paper. The propagation equation of the scattered waves can be 
interpreted as a 
Schroedinger equation describing the nonrelativistic scattering off an 
external potential,
 which is composed of a Newtonian part $\sim (1/r) $ and a dipole part 
 $\sim (GJ/r)\nabla$ (see expression (3) and references \cite{EG,BG}). 
The criterion \cite{JJ} of applicability of the 
Born approximation, which has been adopted in order to obtain equation 
(\ref{eq:helAmpli}), is known to fail for scattering off the Coulomb potential 
although the resulting amplitude turns out to be the exact one \cite{LL} 
(up to a phase).
The remaining part $\sim (GJ/r)\nabla$ of the potential yields,  
for scattering at small angles, the condition 
\beq\label{eq:bornCond}
GJ\omega^2\theta \ll 1
\eeq
On the other hand diffraction effects are negligible, and 
expression (\ref{eq:defImpact}) is sensible, only if the condition
\beq\label{eq:orbitCond}
\omega\, b(\theta)\, \theta \gg 1
\eeq 
is met. When both (\ref{eq:bornCond}) and (\ref{eq:orbitCond}) are satisfied, 
i.e.
\beq\label{eq:constrOmega}
\frac{1}{4GM}\equiv\omega_{\rm min}
\ll \omega \ll \omega_{\rm min}\sqrt{\frac{4bM}{J}}
\equiv\omega_{\rm max}(b)
\eeq 
equation (\ref{eq:deltaTheta}) describes two kinds of
corrections to the classical deflection computed by means of null geodesics.
A distant object emitting unpolarized radiation of the kinds discussed here 
should form, if a rotating body lies between the object and the observation 
point, distinct images for each kind of radiation: for waves of helicity 
$\pm\eta$, equation (\ref{eq:deltaTheta}) predicts two images of equal 
intensity with an angular separation given by
\beq\label{eq:splitAngle} 
\Delta\theta_\eta(b)\simeq 2\eta\omega \frac{J}{M}
\theta_0^3(b)|\ln \theta_0(b)|
\eeq
and centered around the deflection angle 
\beq\label{eq:diffAngle}
\theta_\eta(b)\simeq
\theta_0(b)\left\{1-\frac{\eta}{4}\theta_0^2(b)| \ln \theta_0(b)| \right\}
\eeq

In the case of the Sun, with an impact parameter $b\simeq 1.5\times 10^{9}\,$m 
(twice the radius of the Sun) and with 
$J\simeq1.63\times 10^{41}\,{\rm kg\,m^2\,s^{-1}}$ \cite{Pijp}, one finds 
$\theta_0\simeq 4\times 10^{-6}$, $\omega_{\rm min}=5\times 10^4\,$Hz and 
$\omega_{\rm max}=2.35\times 10^{8}\,$Hz, 
so that the differential angular displacement between the helicity $\eta$ 
and the ``geodesic'' ($\eta=0$) images of the distant object turns out to be
\beq\label{eq:orbSplit}
\frac{\theta_{0}-\theta_{\eta}}{\theta_0}
\simeq  5\, \eta\times 10^{-11}
\eeq
while the differential displacement between the images of the two chiral 
components is
\beq\label{eq:helSplit}
\frac{\Delta\theta_\eta}{\theta_0}\simeq 3.6 \,\eta\,
\omega({\rm Hz})\,\times 10^{-16}
\eeq
In the practical case of Earth based observers, the angular momentum vector 
of the Sun will not be directed as the incoming radiation and equation 
(\ref{eq:defImpact}) cannot apply because the cross 
section (\ref{eq:helAmpli}) 
gets a nontrivial azimuthal dependence. This dependence can be integrated over 
in order to obtain an average deflection angle at a given impact parameter; 
in this case, 
equations (\ref{eq:splitAngle}) and (\ref{eq:diffAngle}) become respectively
\beq\label{eq:avSplitAngle} 
\langle \Delta\theta_\eta\rangle(b)\simeq 2\eta\omega \frac{J\cos\phi}{M}
\theta_0^3(b)|\ln \theta_0(b)|
\eeq
and 
\beq\label{eq:avDiffAngle}
\langle \theta_\eta\rangle (b)\simeq
\theta_0(b)\left\{1+\left[\left(\frac{\omega J\sin\phi}{M}\right)^2 
- \frac{\eta}{4}\right]
\theta_0^2(b)|\ln \theta_0(b)|\right\}
\eeq
where $\cos \phi=\vec{J}\cdot\vec{k}/(J\omega)$.

Equations (\ref{eq:avSplitAngle}) and (\ref{eq:avDiffAngle}) are not 
guaranteed to work 
for $\omega \gtrsim \omega_{\rm max}$, but assuming they do, the null experiment of Harwit et al. 
\cite{Harwit} can be 
given a quantitive description.  In fact, Harwit et al.\@ studied 
electromagnetic radiation $(\eta=1)$ with $\omega \simeq 50\,$GHz 
and $b\simeq 5.5 \times 10^{9}\,$m, leading to 
$\omega_{\rm max}\simeq 4.5 \times 10^{8}\,$ Hz. 
According to the Born approximation result (72), the expected effect is 
\beq\label{eq:previ}
\frac{\Delta\theta_{\eta=1}}{\theta_0}\simeq 1.5\cos\phi\times 10^{-6}
\eeq
which is indeed at least two orders of magnitude below the $10^{-4}$ accuracy they had ($ \, |  \cos \phi \, | \lesssim 0.12 \, $). 
The magnitude (\ref{eq:previ}) 
of the expected deviation from the geometric optics approximation is much 
larger than the estimate 
$\Delta\theta_{\eta=1}/\theta_0\lesssim 10^{-18}$ presented in \cite{MashNat}.

For frequencies $\omega \lesssim \omega_{\rm min}$ the orbit approximation 
fails to be reliable because the diffraction effects become important and the 
helicity interaction 
manifests itself  in the helicity asymmetry of equation (\ref{eq:asymExpr}).

\subsection{Scattering at large angles}

The preceding discussion was limited to the near-forward scattering region 
$\theta \sim 0$, where the Born approximation can be trusted. Large 
gravitational deflections are generated by very compact objects with  
size close to the gravitational radius, so that nonlinear GR effects are 
presumably important. One should then resort to more sophisticated methods 
such as wave decomposition into radial and spherical modes \cite{futt}. 
For spherically symmetric spacetimes ($J=0$), the proper perturbation 
expansion parameter  for propagation of modes with orbital angular momentum 
number $\ell$ is given by $GM\omega / \ell$, so that, for large deflections 
(i.e. low values of $ \ell $), the use of the  linearized metric of equation 
(\ref{eq:metric}) 
is justified in the diffraction limit $\omega \ll \omega_{\rm min}$. 
Note that for black holes this inequality automatically enforces the validity 
of relation (\ref{eq:bornCond}) for $\theta = O(1)$. 

In the diffraction limit, equations (\ref{eq:ampliS1/2}), (\ref{eq:ampliS1}),
(\ref{eq:ampliS2}) can thus be trusted for all values 
of $\theta$. We see in particular that for $ \eta \neq 0$ the backward cross 
section for helicity conserving scattering vanishes (no  glory 
effect) when the incident wave 
vector is aligned with the angular momentum of the deflecting body. 
This is also the case for high frequency scattering, which is best 
described by the JWKB approximation \cite{ma7,matz}. Note that for 
$ \eta=1,2$ the glory for helicity conserving processes is absent whether 
or not the incident wave vector is aligned to $\vec J $. However, while for 
$\eta= 1/2, 1$ the helicity changing amplitudes vanish, this is not the case 
for $\eta=2$ ---as mentioned before--- and the symmetry argument leading to 
the  absence of glory effect for $\eta \neq 0$ \cite{ma7} cannot apply to 
these processes.   
Indeed, equation (\ref{eq:ampliS2}) shows that the helicity changing glory 
scattering cross section of gravitons is nonvanishing, in agreement with the 
analysis of the scattering from Schwarzschild and Kerr black holes 
\cite{futt}. 

\begin{ack}
We wish to thank S. Degl'Innocenti and S. Shore for useful discussions. 
\end{ack}

\end{document}